\documentclass[9pt,twocolumn,twoside]{opticajnl}
\journal{opticajournal}
\setboolean{shortarticle}{true}
\usepackage{graphicx}
\usepackage{subcaption}
\usepackage{lineno}
\usepackage[normalem]{ulem}

\title{Photonic Time-Delayed Quantum Extreme Learning Machine}

\author[1]{Ekaterina Protsenko}
\author[1]{Caterina Vigliar}
\author[1,*]{Francesco Da Ros}

\affil[1]{Department of Electrical and Photonics Engineering, Technical University of Denmark, Kongens Lyngby 2800, Denmark}

\affil[*]{Corresponding author: fdro@dtu.dk}

\begin{abstract}
We propose a photonic quantum extreme learning machine (QELM) based on non-linear interactions in a time-delay non-linear interferometer with photon-number-resolving detection. A silicon spiral waveguide functions simultaneously as a two-mode-squeezing source and delayed feedback loop, eliminating explicit quantum-state preparation and substantially simplifying the experimental  architecture. 

We theoretically demonstrate the QELM's computational capability through binary classification of the non-linear two moons benchmark. The Fock-state probability distribution across temporal modes provides a high-dimensional output feature space. Analysis of the learned output weights and alternative quantum readouts shows that compact feature sets based on photon-number correlations preserve essentially the same classification accuracy while potentially reducing measurement complexity. 

We show that in simulations with realistic optical losses, the QELM maintains its classification accuracy, while the learned weights become distributed across a larger set of features. These results demonstrate that the lossy QELM retains a useful computational feature space and highlight the potential of the time-delay photonic implementation as an experimentally accessible QELM platform.

\end{abstract}

\setboolean{displaycopyright}{false} 

\begin{document}

\maketitle

\subsection{Introduction}
Quantum extreme learning machines (QELMs) exploit the rich dynamics of quantum systems while avoiding the complex training required by general quantum machine-learning models \cite{fujii_harnessing_2017}. QELMs perform nonlinear preprocessing in a fixed quantum system (substrate), while only the linear output layer is optimized. An important property of quantum substrates is their ability to generate high-dimensional output spaces from relatively small physical systems \cite{mujal_opportunities_2021, nerenberg_photon_2024}. Optical quantum states provide diverse degrees of freedom for information encoding and processing, and photonic QELMs have been explored in both continuous-variable \cite{nokkala2021gaussian, garcia-beni_squeezing_2024, maier_continuous-variable_2025} and discrete-variable \cite{nerenberg_photon_2024, yin_experimental_2025, montesinos2026benchmarkingquantumextremelearning, brusaschi_quantum_2026} platforms. These implementations use polarization \cite{nerenberg_photon_2024}, phase  \cite{garcia-beni_squeezing_2024}, or quadrature displacement \cite{maier_continuous-variable_2025} for encoding, with substrate dynamics generated by interferometer meshes \cite{maier_continuous-variable_2025, yin_experimental_2025, montesinos2026benchmarkingquantumextremelearning}, multimode-fiber propagation \cite{joly_harnessing_2025}, or optical non-linearities \cite{garcia-beni_squeezing_2024, brusaschi_quantum_2026}.

In this work, we explore the low-squeezing regime accessible in long silicon waveguides as a resource for QELM implementation. We propose a photonic QELM architecture based on a time-delayed non-linear interferometer (NLI) \cite{ono_observation_2019}, in which the interference between successive temporal instances of two-mode squeezing (TMS) generates time bin encoded squeezed states, and the output is measured using photon number resolving (PNR) detection. The time-delayed scheme reuses a single non-linear element for successive input features, reducing the number of physical components at the cost of sequential processing. We investigate whether this experimentally relevant photonic circuit is sufficient to map the input into a non-linear, high-dimensional discrete feature space for classifying non-linearly separable data and compare it to a feed-forward neural network (FF-NN) and classical signoid Extreme Learning Machines (ELM) implementations. We numerically evaluate the architecture on the two moons task and analyze the contributions of individual Fock-state features to identify reduced readout sets that preserve classification accuracy while potentially lowering measurement complexity. Finally, we assess the accuracy degradation of the QELM under optical loss.

\begin{figure*}[ht]
\centering
\includegraphics[width=\linewidth]{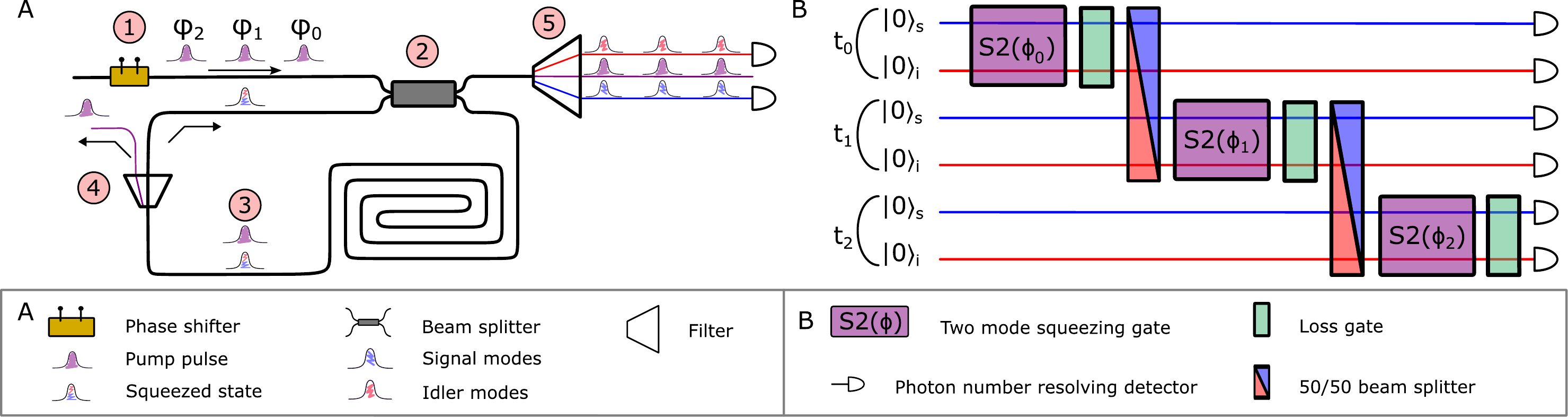}

\caption{
    \textbf{A.} Schematic of the proposed silicon-photonic QELM architecture.
    \textbf{B.} Equivalent quantum circuit used for simulation.
}
\label{fig:scheme}

\phantomsubcaption
\label{fig:schemeA}
\phantomsubcaption
\label{fig:schemeB}

\end{figure*}

\subsection{Architecture}

The proposed architecture is shown in Fig.~\ref{fig:schemeA}. As a proof of concept, we consider the standard two-moons classification benchmark, comprising two input features, $x_1$ and $x_2$, and two non-linearly separable crescent-shaped classes. The input coordinates are encoded in the relative phases of consecutive pulses from a strong mode-locked laser using fast phase-shifter modulation (stage~1). The laser pumps a silicon spiral waveguide whose round-trip length is matched to the pulse repetition period. At stage~2, a 50/50 beam splitter (BS) divides the pump between the measurement path and the spiral-waveguide feedback loop. During the initial time bin $t_0$, spontaneous four-wave mixing in the silicon $\chi^{(3)}$ medium generates a TMS state at signal and idler frequencies $\omega_s$ and $\omega_i$, satisfying $2\omega_p=\omega_s+\omega_i$, where $\omega_p$ is the pump frequency (stage~3). After pump filtering at stage~4, the generated state recirculates and overlaps with the subsequent pump pulse, seeding the next non-linear interaction, which contains both spontaneous and stimulated contributions. At stage~5, the signal and idler modes are separated and measured using PNR detection. We first consider an ideal lossless implementation and subsequently include optical loss. In each time bin, the non-linear waveguide is represented by a TMS gate whose phase is determined by the encoded pump phase. TMS processes generated in successive time bins interfere during subsequent non-linear interactions and are partially coupled into the measurement path, making the output photon-number distribution dependent on their relative phases. The corresponding probabilities are estimated from repeated PNR measurements and used as input features for the logistic-regression readout.

The architecture is simulated by unfolding the time-domain circuit into the path domain, as shown in Fig.~\ref{fig:schemeB}. The modes are arranged as signal--idler pairs for successive time bins. For normalized inputs $x_i^{\mathrm{norm}}\in[0,\pi/2]$, the circuit comprises one reference bin and two input-encoding bins. The pump-phase encoding is defined as $\phi_{p_0}=0$ and $\phi_{p_i}=x_i^{\mathrm{norm}}$, where $\phi_{p_0}$ is the reference phase and $\phi_{p_i}$ corresponds to the input feature $x_i$. In the weak-squeezing regime, we neglect non-linear phase shifts and adopt the undepleted classical-pump approximation. The TMS process is characterized by the phase $\phi=2\phi_p$, up to a constant offset, and the squeezing strength $r$:

\begin{equation}
\hat{\mathrm{U}}_{\mathrm{TMS}}
=
\exp\left[
r e^{i\phi}\hat{a}^{\dagger}_s\hat{a}^{\dagger}_i
-
r e^{-i\phi}\hat{a}_s\hat{a}_i
\right],
\label{eq:TMS}
\end{equation}

where $\hat{a}^{\dagger}_{s,i}$ and $\hat{a}_{s,i}$ are the creation and annihilation operators for the signal and idler modes, respectively. The complex squeezing parameter is $r e^{i\phi}=g|A_p|^2e^{2i\phi_p}$, where $A_p$ is the complex pump amplitude and $g$ is a real gain parameter determined by the waveguide geometry, material properties, and interaction length. We consider $r=0.2$ ($-1.74\,\mathrm{dB}$), corresponding to a single-pair generation probability of approximately $3.7\%$, consistent with reported silicon spiral-waveguide sources \cite{wang_progress_2024}. Propagation-loss gates are omitted in the initial lossless simulation. Consecutive time bins are coupled by separate 50/50 BS acting on the signal and idler modes separately, producing a six-mode system comprising three signal and three idler modes. The BS gate transformation is

\begin{equation}
\hat{\mathrm{U}}_{\mathrm{BS}}=\exp\left[\frac{\pi}{4}\left(e^{i\pi/2}\hat{a}_1\hat{a}_2^\dagger-e^{-i\pi/2}\hat{a}_1^\dagger\hat{a}_2
\right)
\right].
\label{eq:H_ev}
\end{equation}

Numerical simulations were performed using the Strawberry Fields Gaussian backend \cite{killoran_strawberry_2019}. The resulting six-mode continuous-variable state was projected onto a truncated Fock basis to obtain the PNR distribution. Each mode was restricted to at most two photons, consistent with the weak-squeezing regime. The retained probability was $\sum_{\mathbf{n}\in\{0,1,2\}^{M}} p(\mathbf{n}) = 0.9982$, where $\mathbf{n}=(n_1,\ldots,n_M)$ denotes a photon-number configuration over $M=6$ optical modes, confirming the validity of the truncation. We also evaluated experimentally accessible observables, including marginal probabilities, mean and heralded photon numbers, coincidences, and $g^{(2)}$-correlations.

\begin{figure}[t]
\centering
    \includegraphics[width=1\linewidth]{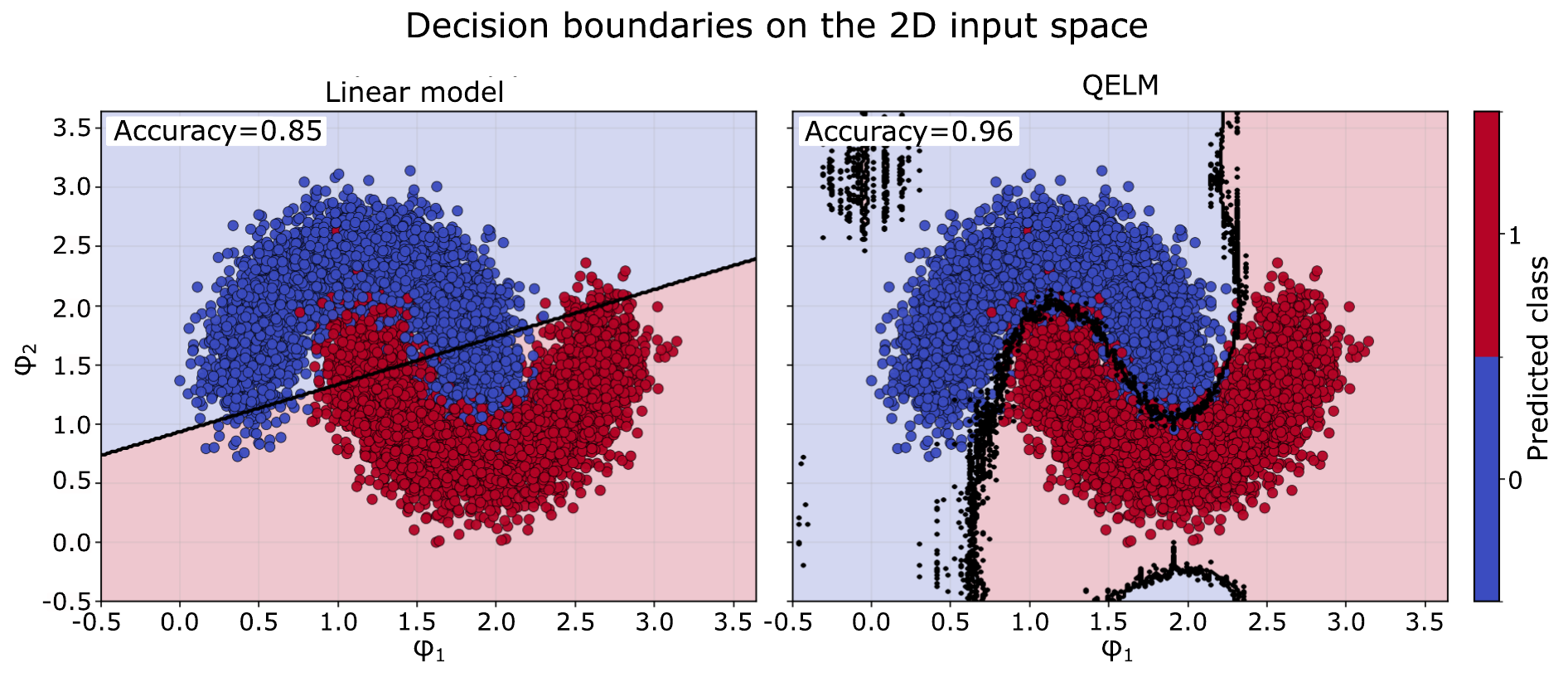}
    \caption{Decision boundaries for the Two Moons classification task obtained using a linear classifier without (left) and with (right) the quantum system. The classical model is trained directly on the encoded phases, whereas the QELM uses probabilities of the truncated Fock states as input features. The labels indicate the corresponding test accuracies.
}
    \label{fig:decision_boundary}
\end{figure}

The two moons classification benchmark is implemented with standard deviation $\sigma=0.2$. The linear model is implemented with \texttt{LogisticRegressionCV} from \texttt{scikit-learn}, where the regularization strength $\lambda$ is selected by 5-fold cross-validation over the grid $\lambda \in \mathrm{logspace}(-12,-2,12)$, with a maximum of 300 iterations. A sweep over dataset size showed that 8000 training samples minimized both training and test errors, while 200 test samples provided stable accuracy estimates with low variance. Figure~\ref{fig:decision_boundary} compares the logistic-regression classification of the two moons dataset with and without the quantum substrate. When trained directly on the encoded phases, the linear classifier achieves a test accuracy of 0.85 and produces a linear decision boundary. Using the quantum substrate as a preprocessing layer increases the test accuracy to 0.96 and yields a non-linear decision boundary, demonstrating a useful high-dimensional nonlinear mapping of the proposed QELM.
 
To assess the statistical robustness all models are trained and evaluated on 100 independently generated datasets. The linear baseline achieved accuracy of $0.8688 \pm 0.023$. The theoretical accuracy limit is estimated using the Bayes-optimal classifier described in Ref.~\cite{metzner_classification_2022}. For a dataset noise level of $\sigma=0.2$, the estimated Bayes-optimal accuracy of $A_{\mathrm{Bayes}}=0.9710$.

To identify output features of the the substrate that contribute most strongly to the QELM accuracy, we apply magnitude-based feature pruning. Specifically, we analyze the logistic-regression weights and define the importance of feature $i$ as the absolute value of its normalized weight, $f_{\mathrm{imp},i}=\left|W_i^{\mathrm{norm}}\right|\in[0,1]$.
Features with importance values below a selected threshold, $W_{\mathrm{cutoff}}$, are removed. The model is then retrained using the retained features, and the test accuracy is evaluated. In total, we analyze 20 levels of weights, with level 1 corresponding to $W_{cutoff}=0$ and level 20 corresponding to $W_{cutoff}=1$. This procedure enables the performance of different observable sets to be compared as a function of the number of retained features.

\begin{figure}
    \centering
    \includegraphics[width=1\linewidth]{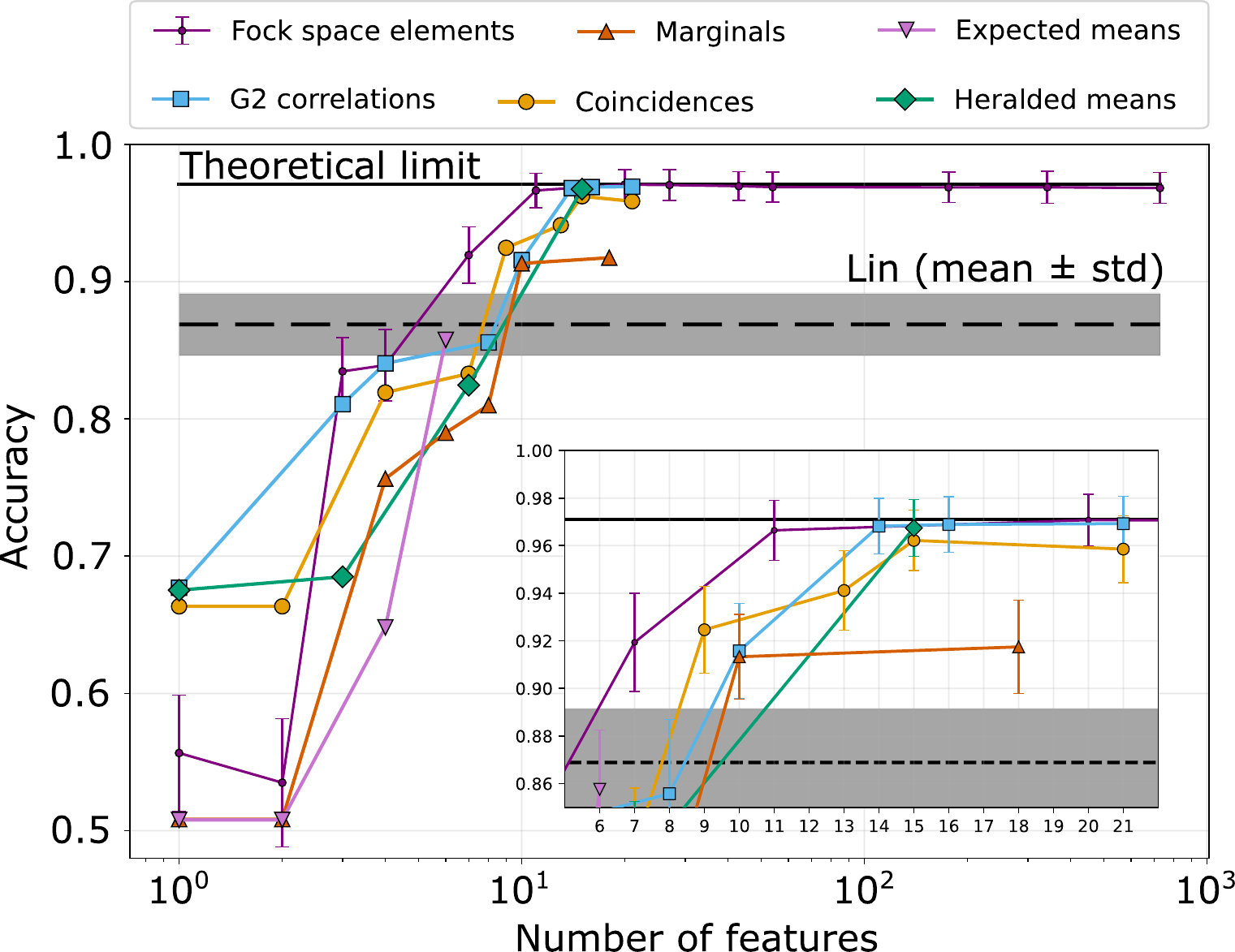}
    \caption{Test classification accuracy versus the number of selected features for different observables. Features are ranked by mean absolute weight and progressively included by relaxing the weight cutoff. Curves show the mean accuracy over 100 realizations; shaded regions and error bars denote one standard deviation. The inset highlights the region between the linear baseline and the theoretical accuracy limit.
}
    \label{fig:num_features}
\end{figure}

The performance of the lossless case using Fock-space probabilities as substrate-output features is shown by the purple curve in Fig.~\ref{fig:num_features}, which presents the mean test accuracy and standard deviation as a function of the number of retained features. Note that these results use exact output probabilities and neglect finite measurement statistics. Using the complete Fock distribution, the model achieves an accuracy of $0.9682 \pm 0.011$. The weight-based feature analysis shows that the QELM surpasses the linear baseline using only seven features and reaches near-optimal performance with 11 features, yielding an accuracy of $0.9665 \pm 0.013$. The highest mean accuracy, $0.9707 \pm 0.011$, is obtained with 20 features. Because these results agree within one standard deviation, 11 features are sufficient to reproduce the optimal performance within statistical uncertainty. The remaining Fock-state probabilities are therefore redundant for this task, indicating that the experimental readout could be simplified. 
For comparison, FF-NNs trained on the same realizations reach $0.9678 \pm 0.018$ with 16 tanh nodes (65 parameters) and $0.9695 \pm 0.012$ with 32 tanh nodes (129 parameters), matching the 11- and 20-feature QELM accuracies, respectively, with comparable spread. The QELM reaches this accuracy by training only a linear readout of N\_features + 1 parameters (21 after feature selection). A 6-node sigmoid ELM reaches $0.9248 \pm 0.0367$, below the 11-feature QELM, while a 20-node yield $0.9706 \pm 0.0112$, matching the 20-feature QELM. Thus, the QELM outperforms a classical ELM whose node count equals the number of quantum modes, but matches it when the node count equals the number of features retained after selection.

\subsection{Comparison of different quantum observables}
Different choices of substrate observables involve different measurement complexity and may lead to different classification accuracies. We therefore investigate whether simpler single-mode observables are sufficient for this task. The marginal probabilities correspond to the probability of measuring a given Fock state in a single mode irrespective of the states of the remaining modes and are defined as $p_i(n) = p(n_i=n)$. Using marginal probabilities as features results in the best classification accuracy ($0.9175 \pm 0.020$) more than 5\% lower than the theoretical maximum, although they still outperform the linear baseline. We also consider the expected means in each mode,
\begin{equation}
\langle n_i \rangle = \sum_{n=1}^{n_{\max}} n p_i(n),
\label{eq:exp_means}
\end{equation}
which provides a compact description of the output state. However, expected means provide test accuracy ($0.8575 \pm 0.025$) worse than marginal probabilities and do not exceed the linear model.

Training on two-mode observables yields significantly better performance. We investigate three classes of observables that capture correlations between pairs of modes. The first is the heralded mean photon number, defined as the expected photon number in one mode conditioned on the presence of at least one photon in another mode:
\begin{equation}
\langle n_i \rangle_{|n_j \geq 1}
= \sum_{n=0}^{n_{\max}} n  p(n_i = n \mid n_j \geq 1).
\label{eq:heralded_clean}
\end{equation}
Although the heralded means require all available features to reach their maximum performance, even this limited information about a second mode yields an accuracy of $0.9674 \pm 0.012$.
The second observable is the coincidence count, i.e. the expected number of photon pairs shared between two modes,
\begin{equation}
C_{ij} = \langle \min(n_i, n_j) \rangle
= \sum_{n=0}^{n_{\max}} n p\left(\min(n_i, n_j) = n\right),
\label{eq}
\end{equation}
which yields a test classification accuracy of $0.9622 \pm 0.013$ using 15 features. Adding more features does not improve the classification accuracy.
The best performance among the two-mode observables is obtained using the second-order correlation function $g^{(2)}$, defined as
\begin{equation}
g_{ij}^{(2)} =
\begin{cases}
\dfrac{\langle n_i n_j \rangle}{\langle n_i \rangle \langle n_j \rangle}, & i \neq j, \\[8pt]
\dfrac{\langle n_i (n_i - 1) \rangle}{\langle n_i \rangle^2}, & i = j.
\end{cases}
\label{eq:g2_clean}
\end{equation}
This observable combines inter-mode correlations with the photon-number statistics of individual modes. Using all 21 features yields a maximum classification accuracy of $0.9693 \pm 0.012$, while a comparable accuracy of $0.9682 \pm 0.012$ is obtained using only 14 features. Overall, the two-mode observables achieve accuracies within one standard deviation of the theoretical maximum with 22 parameters without feature selection. It requires fewer parameters than FF-NNs, outperforms a classical sigmoid ELM with 6 nodes, and achieves slightly lower accuracy than a classical sigmoid ELM with 22 parameters ($0.9709 \pm 0.0107$). This result highlights the choice of observables as an important aspect of QELM design and could mean that higher-order correlations potentially may be useful for more complex tasks.

\subsection{Effect of optical loss}

\begin{figure}[h!]
    \centering
    \includegraphics[width=1\linewidth]{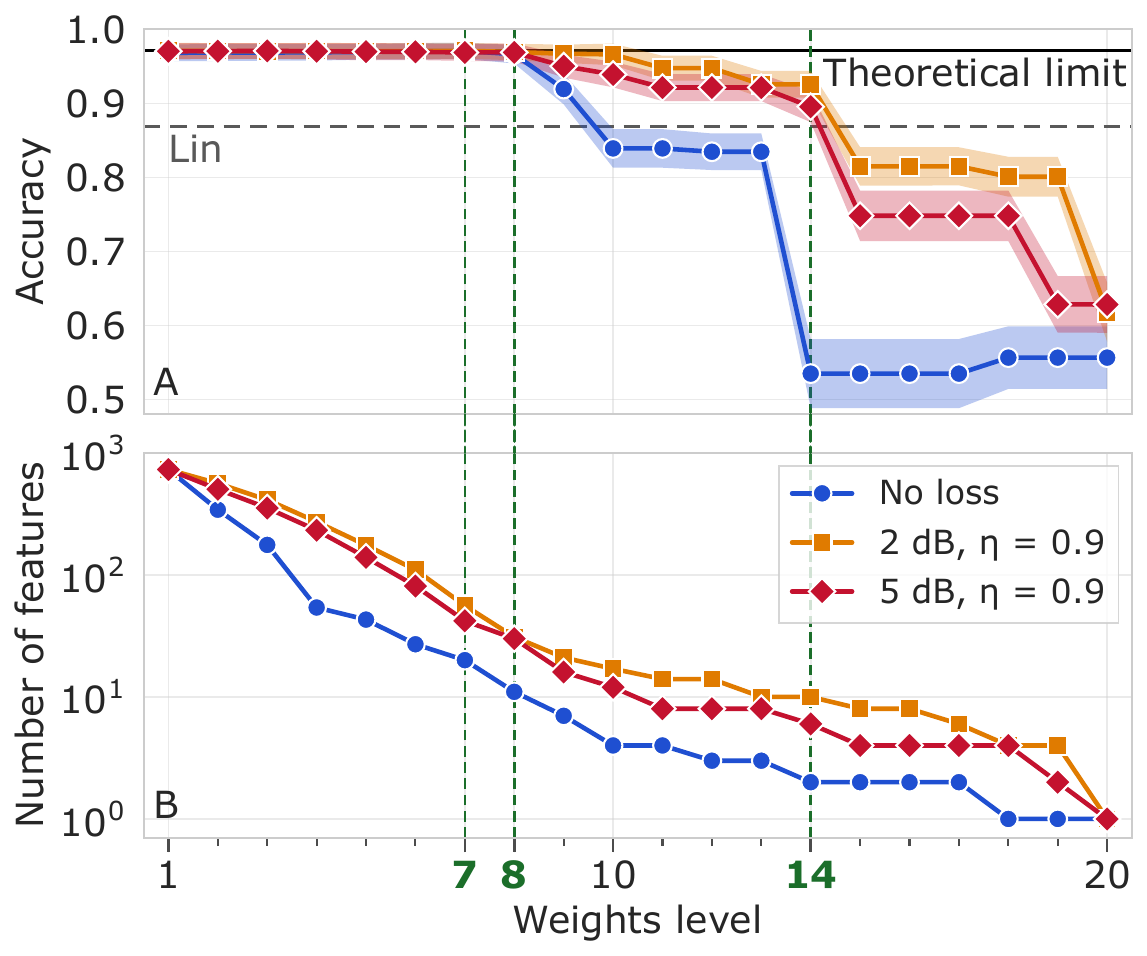}
    \caption{Loss analysis. \textbf{A} Classification accuracy and \textbf{B} number of selected Fock states probabilities as functions of the weights level. The model is trained using only features with $W_i^{\mathrm{norm}} > W_{\mathrm{cutoff}}$ corresponding to each level.
}

\label{fig:loss}

\phantomsubcaption
\label{fig:lossA}
\phantomsubcaption
\label{fig:lossB}
\end{figure}

Optical loss is a major factor limiting the performance of quantum photonic devices \cite{flamini_photonic_2018}. We therefore investigate its effect on the performance of the proposed QELM architecture. We consider loss in the feedback loop and limited detection efficiency as the dominant mechanisms affecting QELM performance, since input coupling losses can be addressed with optimizing the pumping power. For each loss level, the linear readout is retrained on the corresponding lossy features, so the accuracy reflects whether the feature space remains computationally useful. Loss within the feedback loop is modeled by coupling each circulating mode to a vacuum ancilla through a beam-splitter (BS) gate after each S2 operation (Fig. \ref{fig:schemeB}). We consider losses of 2 and 5 dB per circulation, accounting for propagation and filter loss. Finite detection efficiency is modeled by applying a Bernoulli convolution to the ideal photon-number distribution, assuming detection efficiency \(\eta_{\mathrm{det}}=0.9\) in each mode \cite{Zambra2006PhotonStatistics}. Figure~\ref{fig:loss} summarizes the results obtained by applying the weight-based analysis to the Fock-state readout. At weight levels 7 and 8, all models achieve near-optimal accuracy, with the lossless model requiring only 20 and 11 features, respectively, compared with 56 and 31 at 2 dB loss and 42 and 30 at 5 dB. At level 14, the lossy models still outperform the linear baseline with 10 and 6 features, respectively, whereas the lossless model retains only 2 features and drops sharply in accuracy. The number of retained features varies non-monotonically with loss.

\subsection{Conclusion}
We proposed and numerically investigated a silicon-photonic QELM based on the time-delayed interference of TMS processes. The architecture approaches Bayes-optimal performance on the two moons benchmark using experimentally realistic squeezing levels and retains classification accuracy under realistic losses simulation. Analysis of the readout further shows that pairwise correlation observables capture most of the information relevant to the classification task, potentially reducing measurement complexity by identifying substrate readout features that contribute most strongly to the classification and also reducing the number of training parameters. 

The proposed implementation also offers several practical advantages. Information is encoded in the pump phase rather than directly on the generated quantum state, avoiding an additional high-speed modulator in the quantum optical path \cite{ono_observation_2019}.
The squeezed states are generated within the feedback loop, eliminating the need for an external quantum-state source. Reusing the same non-linear element across successive time bins enables different input sizes, although increasing the number of encoded inputs also increases the processing time and the number of modes to be measured. Overall, our results represent a step toward experimentally feasible QELMs implemented on a silicon quantum-photonic platform.

\begin{backmatter}

\bmsection{Acknowledgments} This work is supported by the Danish Research council (QUARCOM grant n. 2032-00161B), the Horizon Europe Research and Innovation Programme (QPIC1550, grant n. 101135785) and the Villum Foundation (OPTIC-AI grant n. VIL29344). We thank R. Zambrini, G.L. Giorgi and M.C. Soriano for helpful discussion.

\smallskip

\bmsection{Disclosures} The authors declare no conflicts of interest.

\bmsection{Data availability} Data underlying the results presented in this
paper are available on a request.

\end{backmatter}

\bibliography{bibliography}

@article{Zambra2006PhotonStatistics,
    author  = {Zambra, Guido and Paris, Matteo G. A.},
    title   = {Reconstruction of photon-number distribution using low-performance photon counters},
    journal = {Physical Review A},
    volume  = {74},
    number  = {6},
    pages   = {063830},
    year    = {2006},
    doi     = {10.1103/PhysRevA.74.063830}
}

@article{fujii_harnessing_2017,
    title = {Harnessing {Disordered}-{Ensemble} {Quantum} {Dynamics} for {Machine} {Learning}},
    volume = {8},
    url = {https://link.aps.org/doi/10.1103/PhysRevApplied.8.024030},
    doi = {10.1103/PhysRevApplied.8.024030},
    number = {2},
    urldate = {2026-06-08},
    journal = {Physical Review Applied},
    publisher = {American Physical Society},
    author = {Fujii, Keisuke and Nakajima, Kohei},
    month = aug,
    year = {2017},
    pages = {024030},
}

@article{mujal_opportunities_2021,
    title = {Opportunities in {Quantum} {Reservoir} {Computing} and {Extreme} {Learning} {Machines}},
    volume = {4},
    copyright = {© 2021 The Authors. Advanced Quantum Technologies published by Wiley-VCH GmbH},
    issn = {2511-9044},
    url = {https://onlinelibrary.wiley.com/doi/abs/10.1002/qute.202100027},
    doi = {10.1002/qute.202100027},
    language = {en},
    number = {8},
    urldate = {2026-06-08},
    journal = {Advanced Quantum Technologies},
    author = {Mujal, Pere and Martínez-Peña, Rodrigo and Nokkala, Johannes and García-Beni, Jorge and Giorgi, Gian Luca and Soriano, Miguel C. and Zambrini, Roberta},
    year = {2021},
    pages = {2100027},
}

@article{nokkala2021gaussian,
  title={Gaussian states of continuous-variable quantum systems provide universal and versatile reservoir computing},
  author={Nokkala, Johannes and Mart{\'\i}nez-Pe{\~n}a, Rodrigo and Giorgi, Gian Luca and Parigi, Valentina and Soriano, Miguel C and Zambrini, Roberta},
  journal={Communications Physics},
  volume={4},
  number={1},
  pages={53},
  year={2021},
  publisher={Nature Publishing Group UK London}
}

@article{maier_continuous-variable_2025,
    author  = {Maier, Benedikt and Spannowsky, Michael and Williams, Simon},
    title   = {Continuous-variable photonic quantum extreme learning machines for fast collider-data selection},
    journal = {EPJ Quantum Technology},
    volume  = {13},
    number  = {1},
    pages   = {70},
    year    = {2026},
    doi     = {10.1140/epjqt/s40507-026-00507-w},
    url     = {https://doi.org/10.1140/epjqt/s40507-026-00507-w}
}

@article{garcia-beni_squeezing_2024,
    title = {Squeezing as a resource for time series processing in quantum reservoir computing},
    volume = {32},
    copyright = {© 2024 Optica Publishing Group},
    issn = {1094-4087},
    url = {https://opg.optica.org/oe/abstract.cfm?uri=oe-32-4-6733},
    doi = {10.1364/OE.507684},
    language = {EN},
    number = {4},
    urldate = {2026-06-08},
    journal = {Optics Express},
    publisher = {Optica Publishing Group},
    author = {García-Beni, Jorge and Giorgi, Gian Luca and Soriano, Miguel C. and Zambrini, Roberta},
    month = feb,
    year = {2024},
    pages = {6733--6747},
}

@article{nerenberg_photon_2024,
author = {Nerenberg, Sam and Neill, Oliver and Marcucci, Giulia and Faccio, Daniele},
year = {2025},
month = {04},
pages = {201-210},
title = {Photon number-resolving quantum reservoir computing},
volume = {3},
journal = {Optica Quantum},
doi = {10.1364/OPTICAQ.553294}
}

@misc{joly_harnessing_2025,
    author = {Joly, Malo and Makowski, Adrian and
              Courme, Baptiste and Porstendorfer, Lukas and
              Wilksen, Steffen and Charbon, Edoardo and
              Gies, Christopher and Defienne, Hugo and
              Gigan, Sylvain},
    title = {Harnessing Photon Indistinguishability in
             Quantum Extreme Learning Machines},
    howpublished = {arXiv:2505.11238},
    year = {2025},
    eprint = {2505.11238},
    archivePrefix = {arXiv},
    primaryClass = {quant-ph}
}

@article{yin_experimental_2025,
    title = {Experimental quantum-enhanced kernel-based machine learning on a photonic processor},
    volume = {19},
    copyright = {2025 The Author(s)},
    issn = {1749-4893},
    url = {https://www.nature.com/articles/s41566-025-01682-5},
    doi = {10.1038/s41566-025-01682-5},
    language = {en},
    number = {9},
    urldate = {2026-06-08},
    journal = {Nature Photonics},
    publisher = {Nature Publishing Group},
    author = {Yin, Zhenghao and Agresti, Iris and de Felice, Giovanni and Brown, Douglas and Toumi, Alexis and Pentangelo, Ciro and Piacentini, Simone and Crespi, Andrea and Ceccarelli, Francesco and Osellame, Roberto and Coecke, Bob and Walther, Philip},
    month = sep,
    year = {2025},
    pages = {1020--1027},
}

@article{ono_observation_2019,
    title = {Observation of nonlinear interference on a silicon photonic chip},
    volume = {44},
    copyright = {© 2019 Optical Society of America},
    issn = {1539-4794},
    url = {https://opg.optica.org/ol/abstract.cfm?uri=ol-44-5-1277},
    doi = {10.1364/OL.44.001277},
    language = {EN},
    number = {5},
    urldate = {2026-06-08},
    journal = {Optics Letters},
    publisher = {Optica Publishing Group},
    author = {Ono, Takafumi and Sinclair, Gary F. and Bonneau, Damien and Thompson, Mark G. and Matthews, Jonathan C. F. and Rarity, John G.},
    month = mar,
    year = {2019},
    pages = {1277--1280},
}

@article{wang_progress_2024,
    title = {Progress on {Chip}-{Based} {Spontaneous} {Four}-{Wave} {Mixing} {Quantum} {Light} {Sources}},
    volume = {5},
    doi = {10.34133/adi.0032},
    journal = {Advanced Devices \& Instrumentation},
    author = {Wang, Haoyang and Zeng, Qiang and Ma, Haiqiang and Yuan, Zhiliang},
    month = jan,
    year = {2024},
}

@article{killoran_strawberry_2019,
    title = {Strawberry {Fields}: {A} {Software} {Platform} for {Photonic} {Quantum} {Computing}},
    volume = {3},
    shorttitle = {Strawberry {Fields}},
    url = {https://quantum-journal.org/papers/q-2019-03-11-129/},
    doi = {10.22331/q-2019-03-11-129},
    language = {en-GB},
    urldate = {2026-06-08},
    journal = {Quantum},
    publisher = {Verein zur Förderung des Open Access Publizierens in den Quantenwissenschaften},
    author = {Killoran, Nathan and Izaac, Josh and Quesada, Nicolás and Bergholm, Ville and Amy, Matthew and Weedbrook, Christian},
    month = mar,
    year = {2019},
    pages = {129},
}

@article{metzner_classification_2022,
    title = {Classification at the accuracy limit: facing the problem of data ambiguity},
    volume = {12},
    copyright = {2022 The Author(s)},
    issn = {2045-2322},
    shorttitle = {Classification at the accuracy limit},
    url = {https://www.nature.com/articles/s41598-022-26498-z},
    doi = {10.1038/s41598-022-26498-z},
    language = {en},
    number = {1},
    urldate = {2026-06-08},
    journal = {Scientific Reports},
    publisher = {Nature Publishing Group},
    author = {Metzner, Claus and Schilling, Achim and Traxdorf, Maximilian and Tziridis, Konstantin and Maier, Andreas and Schulze, Holger and Krauss, Patrick},
    month = dec,
    year = {2022},
    pages = {22121},
}

@article{flamini_photonic_2018,
    title = {Photonic quantum information processing: a review},
    volume = {82},
    issn = {0034-4885},
    shorttitle = {Photonic quantum information processing},
    url = {https://doi.org/10.1088/1361-6633/aad5b2},
    doi = {10.1088/1361-6633/aad5b2},
    language = {en},
    number = {1},
    urldate = {2026-06-08},
    journal = {Reports on Progress in Physics},
    publisher = {IOP Publishing},
    author = {Flamini, Fulvio and Spagnolo, Nicolò and Sciarrino, Fabio},
    month = nov,
    year = {2018},
    pages = {016001},
}

@misc{montesinos2026benchmarkingquantumextremelearning,
    author = {Daniel Montesinos and Gian Luca Giorgi and Roberta Zambrini},
    title = {Benchmarking Quantum Extreme Learning based on Gaussian Boson Sampling},
    howpublished = {arXiv:2606.15230},
    year = {2026}
}

@misc{brusaschi_quantum_2026,
    author = {Brusaschi, Emanuele and Clementi, Marco and Liscidini, Marco and Bajoni, Daniele and Galli, Matteo and Borghi, Massimo},
    title = {Quantum inference on a classically trained quantum extreme learning machine},
    howpublished = {arXiv:2603.20167},
    year = {2026},
    eprint = {2603.20167},
    archivePrefix = {arXiv},
    primaryClass = {quant-ph},
    url = {https://arxiv.org/abs/2603.20167}
}
\clearpage
\onecolumn
\end{document}